\documentclass[twocolumn]{aastex7}

\usepackage{amsmath}
\usepackage{float}
\usepackage{subcaption}
\usepackage{natbib}
\begin{document}

\title{Large Language Models for Limited Noisy Data: A Gravitational Wave Identification Study}

\author[orcid=0009-0002-0541-7182]{Yixuan Li}
\affil{School of Mathematics and Physics, University of South China, Hengyang, 421001, China}
\affil{ICRANet-AI, Brickell Avenue 701, Miami, FL 33131, USA}
\email{} 

\author[0009-0002-7655-6737]{Yuhao Lu}
\affil{School of Computer Science, University of South China, Hengyang, 421001, China}
\affil{ICRANet-AI, Brickell Avenue 701, Miami, FL 33131, USA}
\email{} 

\author[0009-0002-9188-5029]{Yang Liu}
\affil{Department of Physics E. Pancini, University Federico II, Naples 80126, Italy}
\affil{ICRANet-AI, Brickell Avenue 701, Miami, FL 33131, USA}
\email{} 

\author[0000-0002-1343-3089]{Liang~Li}
\affiliation{Institute of Fundamental Physics and Quantum Technology, Ningbo University, Ningbo, Zhejiang 315211, People's Republic of China}
\affiliation{School of Physical Science and Technology, Ningbo University, Ningbo, Zhejiang 315211, People's Republic of China}
\email{}

\author[0000-0003-0829-8318]{R.~Ruffini}
\affil{ICRANet, Piazza della Repubblica 10, 65122 Pescara, Italy}
\affil{ICRA, Dip. di Fisica, Sapienza Universit\`a  di Roma, Piazzale Aldo Moro 5, I-00185 Roma, Italy}
\affil{ICRANet-AI, Brickell Avenue 701, Miami, FL 33131, USA}
\affil{Universit\'e de Nice Sophia-Antipolis, Grand Ch\^ateau Parc Valrose, Nice, CEDEX 2, France}
\affil{INAF,Viale del Parco Mellini 84, 00136 Rome, Italy}
\email{} 

\author[0000-0003-3010-7661]{Di Li}
\affil{New Cornerstone Science Laboratory, Department of Astronomy, Tsinghua University, Beijing 100084, China}
\affil{National Astronomical Observatories, Chinese Academy of Sciences, Beijing 100101, China}
\affil{State Key Laboratory of Radio Astronomy and Technology, Beijing 100101, China}
\affil{ICRANet, Piazza della Repubblica 10, 65122 Pescara, Italy}
\email{} 

\author[0000-0002-3539-7103]{Rong-Gen~Cai}
\affiliation{Institute of Fundamental Physics and Quantum Technology, Ningbo University, Ningbo, Zhejiang 315211, People's Republic of China}
\email{} 

\author[0009-0008-2830-9866]{Xiaoyan Zhu}
\affiliation{School of Mathematics and Physics, University of South China, Hengyang, 421001, China}
\email{} 

\author[0000-0002-4282-066X]{Wenbin Lin}
\affil{School of Computer Science, University of South China, Hengyang, 421001, China}
\affil{School of Mathematics and Physics, University of South China, Hengyang, 421001, China}
%\affil{School of Nuclear Science and Technology, University of South China, Hengyang, 421001, China.}
\affil{ICRANet, Piazza della Repubblica 10, 65122 Pescara, Italy}
\email{lwb@usc.edu.cn} 

\author[0000-0001-7959-3387]{Yu Wang}
\affil{ICRA, Dip. di Fisica, Sapienza Universit\`a  di Roma, Piazzale Aldo Moro 5, I-00185 Roma, Italy}
\affil{ICRANet, Piazza della Repubblica 10, 65122 Pescara, Italy}
\affil{ICRANet-AI, Brickell Avenue 701, Miami, FL 33131, USA}
\affil{INAF -- Osservatorio Astronomico d'Abruzzo, Via M. Maggini snc, I-64100, Teramo, Italy.}
\email{yu.wang@icranet.org}

\correspondingauthor{Yu Wang (yu.wang@icranet.org), Xiaoyan Zhu (xyzhu0128@163.com), Wenbin Lin (lwb@usc.edu.cn), Di Li (dili@tsinghua.edu.cn)}

\begin{abstract}
This work investigates whether large language models (LLMs) offer advantages over traditional neural networks for astronomical data processing, in regimes with non-Gaussian, non-stationary noise and limited labeled samples. Gravitational wave observations provide a suitable test case, using only 90 LIGO events, finetuned LLMs achieve 97.4\% accuracy for identifying signals. Further experiments show that, in contrast to traditional networks that rely on large simulated datasets, additional simulated samples do not improve LLM performance, while scaling studies reveal predictable gains with increasing model size and dataset size. These results indicate that LLMs can extract discriminative structure directly from observational data and provide an efficient assessment for gravitational wave identification. The same strategy may extend to other astronomical domains with similar noise properties, such as radio or pulsar observations.
\end{abstract}

%\keywords{Gravitational wave signal classification, DeepSeek, LLaMA3, Scaling laws}

\section{Introduction} \label{sec:introduction}

LLMs have become central to recent progress in machine learning, and many proposals in different fields now aim to ``add an LLM'' as a universal solution. In practice, however, our experience suggests that LLMs are not always superior to traditional networks, especially for tasks that require precise numerical prediction \citep{wiggins2022opportunities,kasneci2023chatgpt,chang2024survey}. This raises a concrete scientific question: \emph{for which types of scientific data and purposes are LLMs actually advantageous compared with traditional neural networks?}

The starting point is to consider the way LLMs process data. LLMs operate on discrete patch tokens rather than dense pixel grids, and attention layers compare these tokens globally rather than locally \citep{vaswani2017attention, han2022survey}. As a result, the model emphasizes large-scale structure, long-range coherence, and relationships across the entire input, instead of relying on small convolutional windows and accurate calculations. This inductive bias is naturally suited to data whose discriminative information resides in global morphology rather than in local texture. It is also compatible with datasets in which the noise is non-Gaussian and non-stationary, because global attention tends to suppress isolated transient artifacts while highlighting extended coherent patterns. 

The second indication comes from our previous tests \citep{wang2024aiuniverse, wang2025reflections}, in which we fine-tuned GPT-based models on a diverse collection of astrophysical datasets, including Sloan Digital Sky Survey spectra, gamma-ray burst properties, and black hole spin measurements. A single model could perform multiple tasks without requiring separate specialized networks. Furthermore, even when trained on less than 1\% of the available data and using reduced-resolution inputs, the models needed only one or two epochs of finetuning to reach competitive accuracy. These results suggest that LLMs adapt rapidly in small-sample regimes and can learn high-level structure efficiently, behaviors that differ from those of conventional neural networks.

These considerations point naturally toward gravitational wave data as a test case. In compact binary coalescences, the discriminative information is carried by the global chirp trajectory in the time-frequency plane, a smooth and coherent pattern that depends on the overall evolution rather than on local numerical details. Such morphology aligns well with the token-attention mechanism of LLMs, which emphasizes long-range structure across the entire input. By contrast, the noise in gravitational wave detectors is dominated by non-Gaussian and non-stationary transients \citep{2019PhRvL.122t1101C, 2021PhRvD.104f3034Z, 2022CQGra..39x5013D, 2023ApJ...949L..41L}, including short duration glitches that produce localized high amplitude features. These artifacts often bias convolution based models but should be naturally downweighted by global attention mechanisms that compare all tokens jointly. Moreover, the number of  gravitational wave events remains limited, with roughly 90 detections by the end of the LIGO O3 observing run. 

%Among existing astronomical datasets, gravitational wave observations most clearly satisfy these conditions. The discriminative information in a compact binary signal lies in its global chirp morphology, not local fluctuations. Detector noise is known to be highly non-Gaussian and non-stationary, with frequent transient glitches \citep{2019PhRvL.122t1101C, 2021PhRvD.104f3034Z, 2022CQGra..39x5013D, 2023ApJ...949L..41L}. And only 90 observed events till the LIGO O3 run. 

%Gravitational waves are ripples in spacetime, generated by violent cosmic events such as black hole mergers and supernova explosions. These massive astrophysical disturbances perturb the fabric of spacetime, causing waves that propagate in all directions at the speed of light, carrying critical information about their sources and offering inshights into the nature of gravity. Since the first direct detection by LIGO in 2015 \citep{abbott2016advligo}, gravitational wave astronomy has rapidly developed, enabling a new sensory channel for humans to explore and understand the cosmos. Despite these breakthroughs, 

Reliably identifying gravitational wave signals remains a major challenge, as the signals are often extremely weak and deeply embedded in complex noise \citep{beker2011newtonian}. Matched filtering \citep{1994Performance} is optimal for known signals in stationary Gaussian noise, but detector data contain non-Gaussian and non-stationary transients that produce spurious high-SNR triggers, requiring additional vetoes such as the $\chi^2$ time–frequency test and reweighted statistics \citep{gabbard2018deepgw,2005PhRvD..71f2001A}. In the fourth observing run, the GstLAL compact binary search pipeline uses a template bank constructed via the manifold method, with a minimal match of $\sim$97\% to limit event-rate loss to $\sim 10\%$ \citep{2016CQGra..33u5004U, 2024PhRvD.109d4066S}. This bank covers the parameter space of aligned-spin neutron star and black hole systems up to mass ratio of 20 and includes on the order of hundreds of thousands of templates. Generating and filtering against such a large bank imposes high computational cost and significant latency, especially as waveform durations increase at lower frequencies \citep{2024PhRvD.109d4066S}. These factors reduce real-world detection accuracy and hinder low-latency operation.

Neural networks have been shown to reach sensitivities comparable to matched filtering while offering orders of magnitude reductions in inference time. Prior work using convolutional network (CNN) architectures \citep{george2018deep, gabbard2018deepgw, 2023PhLB..84037850Q} demonstrated that, once trained, these models can classify gravitational wave candidates within milliseconds and reproduce the detection performance of traditional pipelines. These studies typically rely on large collections of simulated compact binary coalescence signals, often numbering from several thousand to several hundred thousand templates, injected into either real or Gaussian noise. While effective, this strategy requires substantial computational effort to generate training sets and is inherently limited by the fidelity of the waveform models. Furthermore, networks trained primarily on simulations may experience domain mismatch when applied to real interferometer data, where non–Gaussian and non–stationary noise artifacts are common. These factors motivate exploring alternative approaches that depend less on massive synthetic datasets and can learn directly from observational data.

LLMs offer a different paradigm. In gravitational wave research, however, they have so far been used only in auxiliary roles, such as providing domain-aware heuristics to guide algorithm design \citep{2025arXiv250803661W}, and, to our knowledge, have not yet been directly applied to the processing and identification of observational data.

This article is structured as follows. Section~\ref{sec:motivation} outlines the motivation for this work. Building on how LLMs process data and on our previous finetuning experience, we hypothesize that LLMs may handle non-Gaussian and non-stationary noise more effectively than traditional neural networks and also adapt well when only limited training data are available. This motivates a direct experimental evaluation. Section~\ref{sec:real-world-eval} presents our observational data experiment, including dataset construction, model configuration, training, and classification results. Section~\ref{sec:simulation-based-eval} extends the analysis by examining the influence of simulated datasets, model size, and dataset size. Section~\ref{sec:other-data} discusses how similar data characteristics arise in other astronomical domains. Finally, Section~\ref{sec:conclusions} summarizes the main findings of this study.

\section{Motivation  \&  Prediction}\label{sec:motivation}

\subsection{Non-Gaussian and Non-Stationary Noise}
Interferometric gravitational wave detectors measure a high sampling rate strain time series,
\begin{equation}
    d(t) = h(t;\theta) + n(t),
\end{equation}
where $h(t;\theta)$ is the astrophysical signal determined by the source parameters $\theta$, and $n(t)$ denotes the instrumental noise.  
Standard analysis \citep{2009LRR....12....2S,2012PhRvD..85l2006A,2021PhRvD.104f3034Z} assume that the noise is a stationary Gaussian process, fully characterized by its one-sided power spectral density (PSD) $S_{n}(f)$:
\begin{equation}
    \langle \tilde{n}(f)\,\tilde{n}^{*}(f') \rangle = 
    \frac{1}{2}\, S_{n}(f)\, \delta(f-f'),
\end{equation}
which enables the construction of the noise-weighted inner product,
\begin{equation}
\langle a, b \rangle 
= 4\,\Re \int_{f_{\min}}^{f_{\max}}
\frac{\tilde{a}(f)\,\tilde{b}^{*}(f)}{S_{n}(f)}\,{\rm d}f,
\end{equation}
and the corresponding Gaussian likelihood,
\begin{equation}
p(d \mid \theta) \propto 
\exp\!\left[
-\frac{1}{2}\,
\langle d - h(\theta),\, d - h(\theta)\rangle
\right].
\end{equation}

Real detector data, however, violate these assumptions in several important ways.  
A dominant feature is the presence of short-duration, high-amplitude transient artifacts known as \emph{glitches}.  
These events appear as localized or narrowband structures in the time–frequency plane and introduce heavy tails in the noise distribution, which is incompatible with Gaussianity.  
Their morphology, duration, and amplitude vary significantly and cannot be described by a fixed PSD.  
The Gravity Spy project\footnote{\url{https://www.zooniverse.org/projects/zooniverse/gravity-spy}} provides a representative catalogue of such transients.

In addition, gravitational wave noise is strongly non-stationary.  
Environmental variations, seismic motion, thermal drifts, and changes in interferometer control states cause the PSD to evolve over minutes to hours.  
This can be captured through a time-dependent PSD, $S_{n}(f;t)$, or a local noise autocorrelation,
\begin{equation}
    C_{n}(\tau; t)
    = \big\langle n(t')\,n(t'+\tau)\big\rangle_{t' \in \text{window around }t},
\end{equation}
both of which vary with time.  
These effects break the assumption that a single stationary PSD can describe long data segments.

In contrast, astrophysical compact-binary signals exhibit coherent and predictable evolution \citep{2009LRR....12....2S}.  
The instantaneous frequency of an inspiral follows
\begin{equation}
    \frac{{\rm d}f}{{\rm d}t} \propto f^{11/3} M_{\rm chirp}^{5/3},
\end{equation}
producing a smooth, monotonic chirp track in the time–frequency plane.  
Glitches, lacking long-range coherence, occupy short temporal or spectral intervals. Even in the seconds long cases, they often appear as broadband noise blocks, irregularly drifting streaks, or piecewise concatenated structures. They can be only long in duration, but they are not coherent. 

The violation of the Gaussian and stationary assumptions has direct implications for signal identification.  In matched-filter searches, the detection statistic relies on the expected distribution of the signal-to-noise ratio under stationary Gaussian noise.  Non-Gaussian transients generate large spurious SNR peaks, inflating the background tail and forcing higher detection thresholds, which in turn obscure low-SNR astrophysical events \citep{2010PhRvD..82j3007L, 2020PhRvD.101d2003V}.  Non-stationary noise further degrades the optimality of matched filtering because the weighting by $S_{n}(f)$ is no longer accurate when the PSD drifts on short timescales.

\subsection{Token-Based LLMs Match the Noise Structure}

To analyze data dominated by non-Gaussian and non-stationary noise, a model must satisfy two requirements:
it must remain robust to localized heavy-tailed noise, and it must be able to extract coherent patterns that extend across time and frequency. Traditional neural networks like CNNs are limited in this setting because their feature extraction relies on localized convolution kernels \citep{2023CmPhy...6..212Z,2025FrPhy..2045301Z}. A typical convolution layer computes \citep{lecun2002gradient, goodfellow2016deep}
\begin{equation}
    z_{k}(u,v) 
    = \sigma\!\left[
    \sum_{i,j} w_{k}(i,j)\, I(u+i, v+j) + b_{k}
    \right],
\end{equation}
where $I(x,y)$ is the input image, $w_{k}$ is the local kernel, $b_{k}$ is the bias,
$(u,v)$ is the output location, and $\sigma(\cdot)$ is a nonlinear activation.
Since each activation is dominated by a small neighborhood of pixels, any strong transient spike inside that neighborhood can overwhelm the representation, causing CNNs to misinterpret glitches as salient features. Meanwhile, the global waveform morphology, distributed across many time–frequency bins, may receive little weight. This effect is stronger for seconds long glitches. A CNN expands its receptive field by stacking layers, and at finite depth it still represents structures on multi second timescales by piecing together local features. If a glitch shows high energy or sharp edges in many local windows, these local responses can accumulate, and the model can easily interpret the result as a signal like pattern. This failure is not because the glitch is too short, but because it is long enough in time to keep activating local convolutional kernels.

LLMs with tokenization \citep{2015arXiv150807909S} and self-attention \citep{vaswani2017attention} adopt a fundamentally different mechanism. The input time–frequency map $X$ is divided into $N$ non-overlapping patches $\{P_i\}$, each of which is mapped into a vector embedding:
\begin{equation}
    x_i = E(P_i), \quad i = 1,\dots, N,
\end{equation}
where $E(\cdot)$ is a learned embedding function. Each token $x_i$ therefore summarizes the overall pattern inside patch $P_i$, rather than the value of any single pixel. Localized noise bursts typically affect only a few pixels inside a patch and therefore only weakly perturb its embedding, while a coherent waveform leaves correlated structure across multiple patches. When a transient spans a longer interval, it can influence many patches, but the embeddings still reflect whether neighboring patches participate in a consistent track or vary irregularly across time and frequency.

Self-attention then forms global relationships across all tokens. For each token $x_i$, a query vector $Q_i$, key vector $K_i$, and value vector $V_i$ are computed \citep{vaswani2017attention}:
\begin{equation}
    Q_i = W_Q x_i, \quad 
    K_i = W_K x_i, \quad 
    V_i = W_V x_i,
\end{equation}
where $W_Q$, $W_K$, and $W_V$ are learned matrices. The attention weights are
\begin{equation}
    \alpha_{ij} 
    = \frac{
        \exp\!\big( Q_i K_j^{\mathrm T} / \sqrt{d} \big)
      }{
        \sum_{\ell=1}^{N}
        \exp\!\big( Q_i K_\ell^{\mathrm T} / \sqrt{d} \big)
      },
\end{equation}
where $d$ is the token dimension. The output token is
\begin{equation}
    y_i = \sum_{j=1}^{N} \alpha_{ij} V_j.
\end{equation}

If the input contains a gravitational wave signal, the patches lying along the chirp trajectory share coherent structure, making their embeddings mutually similar. This yields large dot products $Q_i K_j^{\mathrm T}$ and correspondingly large attention weights $\alpha_{ij}$. Self-attention thus amplifies long-range, physically meaningful correlations across the time–frequency plane. Glitches, by contrast, have weak cross-patch coherence, causing their attention weights to remain localized and preventing them from dominating the global representation.

Unlike CNNs, self-attention does not assume translational invariance, which is particularly important for non-stationary noise. With positional encodings added to each token, $x_i + p_i$, the attention mechanism can explicitly learn how noise statistics vary with time, enabling the model to distinguish slow PSD drifts from physically meaningful signal evolution.

To conclude: patch-level embeddings suppress local heavy-tailed noise, and global attention promotes extended coherent patterns. This structural alignment with the statistical properties of gravitational wave data provides a natural motivation for exploring LLMs as analysis tools in the presence of strong non-Gaussian and non-stationary noise.

\section{Testing LLMs by Gravitational Wave} \label{sec:real-world-eval}

\subsection{Gravitational Wave Dataset Construction}
\label{sec:data-preprocessing}

\begin{figure}
	\centering
%	\begin{subfigure}
		\includegraphics[width=0.48\textwidth]{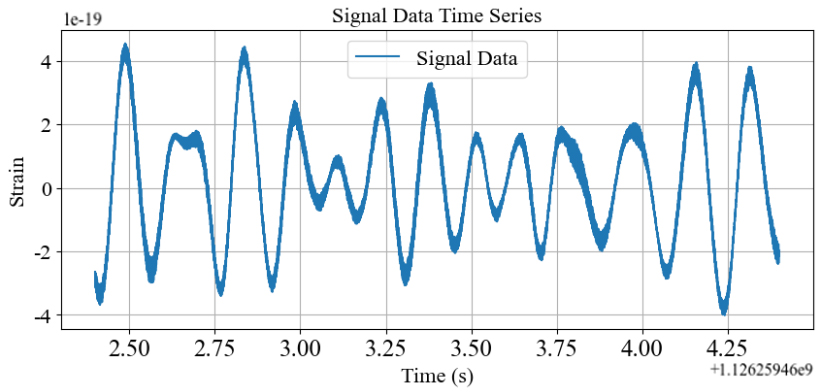}
        \includegraphics[width=0.48\textwidth]{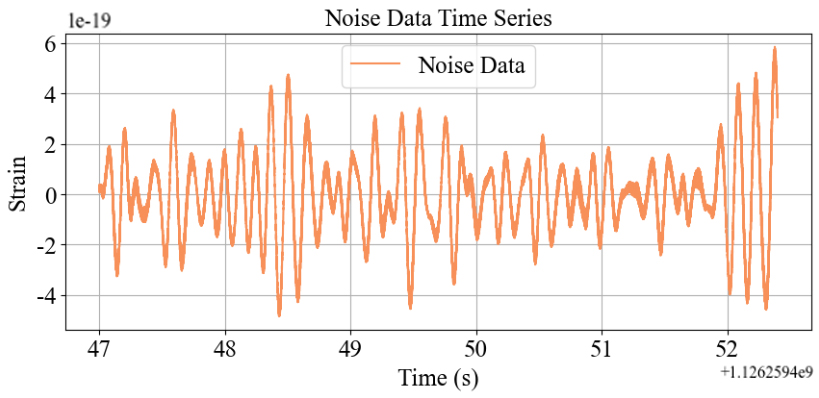}
		\caption{Comparison between a gravitational wave signal segment and a noise only segment. \textbf{Top}: representation of the GW150914 event. \textbf{Bottom}: Representative noise segment from the same detector band.}
        \label{fig:fig1}
\end{figure}

We first construct a dataset based solely on observational data. All events are selected from the publicly available PyCBC\citep{biwer2019pycbc} event catalog using \texttt{pycbc.catalog.Catalog}. The event list follows the GWTC-3 confident event versions released by GWOSC \citep{KAGRA:2021vkt}, which include all compact binary merger events detected by LIGO and Virgo during the O1 (2015), O2 (2016 to 2017), O3a (2019), and O3b (2019 to 2020) observing runs, in total 90 events. For each event, strain time series from the LIGO Hanford (H1) and Livingston (L1) detectors are obtained, together with event-specific metadata such as GPS timestamps and data quality information, using \texttt{pycbc.catalog.Merger}. These timestamps are used to extract signal and noise segments, as shown in Figure~\ref{fig:fig1}. A sliding window of 2 seconds with a 1-second stride is applied to the strain data to generate overlapping segments, ensuring full coverage of the signal interval and increasing the number of training samples. In total, 1728 signal segments are produced.

%Signal labeling is guided by physical event parameters, such as component masses and spins, sourced from the GWOSC database\footnote{\url{https://gwosc.org/eventapi/html/GWTC/}}. 

The duration of each event was estimated from the time it entered the 20 Hz detection band to merger completion using the \texttt{SEOBNRv4\_opt} waveform model\footnote{\url{https://gwosc.org/eventapi/html/GWTC/}}. Segments overlapping with the theoretical event duration are labeled as positive. If the event is shorter than 2 seconds, the segment starting at the event time is labeled; otherwise, all overlapping windows covering the full signal are labeled as positives. Segments outside this duration are treated as negatives.

Due to the significant imbalance between noise and signal segments, we employ an oversampling strategy\citep{he2009learning} to replicate positive samples to reach class parity. This helps mitigate bias toward the majority (noise) class during training.

We further follow the standard gravitational wave analysis procedure \citep{flanagan2005basics}, first to apply a bandpass filter on the raw data to retain frequencies most relevant to signal detection. Specifically, an eighth-order Butterworth filter was used to isolate the 20–500 Hz frequency band, as determined through empirical evaluation. The filter was implemented using the \texttt{SciPy}\citep{virtanen2020scipy} signal processing library, and a Tukey window function was applied prior to filtering to mitigate edge effects. The Butterworth filter is defined as:

\begin{equation}
	H(f) = \frac{1}{\sqrt{1 + \left( \frac{f_c}{f} \right)^{2n} }},
\end{equation}

where \( f_c \) is the cutoff frequency and \( n = 8 \) is the filter order. The Tukey window function used to smooth the signal edges is given by:

\begin{equation}
	w(t) =
	\begin{cases}
		\frac{1}{2} \left[1+\cos \left(\pi \left( \frac{2t}{\alpha T} - 1 \right) \right) \right], & 0 \leq t < \frac{\alpha T}{2} \\
		1, & \frac{\alpha T}{2} \!\leq\! t \leq \!T \!- \!\frac{\alpha T}{2} \\
		\frac{1}{2} \!\left[1\!+\!\cos \left(\pi \!\left( \frac{2t}{\alpha T}\! -\! 2 + \!\frac{2}{\alpha} \right) \right) \right], & T\! - \!\frac{\alpha T}{2} \!\leq \!t \!\leq T
	\end{cases}
\end{equation}

with \( T \) denoting the total signal duration and \( \alpha \) set to 0.2. After filtering, each signal is normalized by its maximum absolute value:

\begin{equation}
	x_{\text{norm}} = \frac{x}{\max(|x|)} ,
\end{equation}

To transform the signal to the time-frequency domain, we apply the Constant-Q Transform (CQT) \citep{brown1991calculation} using the nnAudio toolkit\citep{cheuk2020nnaudio}. It maps the waveform into a 2D time-frequency space with logarithmically spaced frequency bins. The CQT is configured with a sampling rate of 2048 Hz, a frequency range of 20–500 Hz, and a hop length of 64. The CQT output for the \( k \)-th frequency bin is computed as:

\begin{equation}
	X(k) = \sum_{n=0}^{N_k - 1} x(n) \cdot w_k(n) \cdot e^{-2\pi i n k / N_k},
\end{equation}

where \( x(n) \) is the input signal, \( w_k(n) \) is the analysis window, and \( N_k \) is the window length corresponding to bin \( k \).

The resulting 2D time-frequency matrices are then discretized into sequences to be compatible with the input requirements of LLMs. For each sample, up to 64 frame-wise feature vectors are extracted and clustered using the KMeans algorithm \citep{macqueen1967classification}. This quantization converts continuous-valued vectors into discrete integers, forming compact sequences that retain essential information. Integers sequences from the three detectors are concatenated to form a unified input.

The final dataset is serialized in JSONL format, conforming to the prompt-response schema commonly used in ChatGPT-style models. Each record contains sequences from all detectors and a binary label. The dataset is split into training and testing subsets with an 80\%-20\% ratio to ensure reliable model evaluation.

\subsection{Model Architecture}

To identify gravitational wave signals, this study employs \texttt{Meta-Llama-3-8B-Instruct}, a recently released open-weight LLM developed by Meta AI \citep{2024arXiv240721783G}. Llama-3-8B-Instruct is a transformer-based model with 8 billion parameters. This parameter size strikes a balance between accuracy and resource usage as we will confirm later: large enough to capture complex sequential dependencies, and small enough to be deployed on local hardware without prohibitive cost.

The input sequences are first passed through an embedding layer to be tokenization before being processed by the Transformer layers. Each Transformer block consists of a multi-head attention mechanism and a position-wise feedforward network. The model also integrates position encoding via rotational embeddings \citep{su2024roformer}, layer normalization using RMSNorm \citep{zhang2019rmsnorm}, and SwiGLU activations \citep{shazeer2020glu}, with intermediate layer dimensionality set to \( \frac{8}{3} d_{\text{model}} \).
%To adapt the LLaMA3 model to the gravitational wave classification task, we apply low-rank adaptation (LoRA)\citep{hu2022lora}, which inserts trainable low-rank matrices into specific attention layers to enable parameter-efficient finetuning. 
In addition, a binary classification head is appended to the model:

\begin{equation}
	\hat{y} = \sigma(W_h x + b_h),
\end{equation}

where \( W_h \) and \( b_h \) are the weights and bias, and \( \sigma \) denotes the sigmoid function.

\subsection{Model Training}

Finetuning of the \texttt{Meta-Llama-3-8B-Instruct} model is conducted on three RTX 3090 GPUs. To accommodate memory constraints, the batch size was set to 1 per GPU, with gradient accumulation of 8 steps to simulate a larger effective batch size. AdamW optimizer \citep{2017arXiv171105101L} with a relatively high learning rate of $5.0 \times 10^{-4}$ was used to accelerate convergence, combined with a weight decay of $1.0 \times 10^{-3}$ to prevent overfitting. The training process employed a cosine annealing scheduler \citep{2016arXiv160803983L}, which provides smooth and gradual learning rate decay and demonstrated stable performance across experiments.

To reduce training complexity and memory usage, we adopted the Low-Rank Adaptation (LoRA) technique \citep{hu2022lora}. LoRA restricts weight updates to low-rank matrices, enabling efficient adaptation of large models with limited resources. The finetuning configuration included a rank of 8, an alpha scaling factor of 16, and a dropout rate of 0.1. This setup allowed for effective model specialization without significant overfitting, even under constrained batch sizes. 

%Model training uses the AdamW optimizer \citep{2017arXiv171105101L} with a learning rate of \(10^{-5}\).  
The objective function is the binary cross-entropy loss,
\begin{equation}
    \mathcal{L}
    = -\frac{1}{N}\sum_{i=1}^{N}
    \left[
        y_i \log(\hat{y}_i)
        + (1 - y_i)\log\!\left(1 - \hat{y}_i\right)
    \right],
\end{equation}
where \(N\) is the number of training samples, \(y_i\in\{0,1\}\) is the true class label, and \(\hat{y}_i\in[0,1]\) is the predicted probability that the \(i\)-th sample contains a gravitational wave signal.

Model checkpoints were saved every 500 training steps, and validation performance was monitored to ensure stable convergence and avoid overfitting.

\subsection{Results on Observational Data}

We evaluate the performance of LLMs on LIGO observations using only 90 gravitational wave events for finetuning. The models are initialized from their general pretrained states and receive no additional training on simulated gravitational wave signals. Under this setting, the training converges rapidly: the loss decreases sharply within the first few steps and stabilizes by only two epochs, with no indication of overfitting despite the limited size of the dataset.

\begin{figure}
	\centering
	% \begin{subfigure}[b]{0.5\textwidth}
	% 	\includegraphics[width=\textwidth]{fig/fig2-1.png}
	% 	\caption{Training Loss and Validation Accuracy}
	% 	\label{fig:fig2-1}
	% \end{subfigure}
%	\hspace{0.02\textwidth}
%	\begin{subfigure}[b]{0.45\textwidth}
		\includegraphics[width=0.45\textwidth]{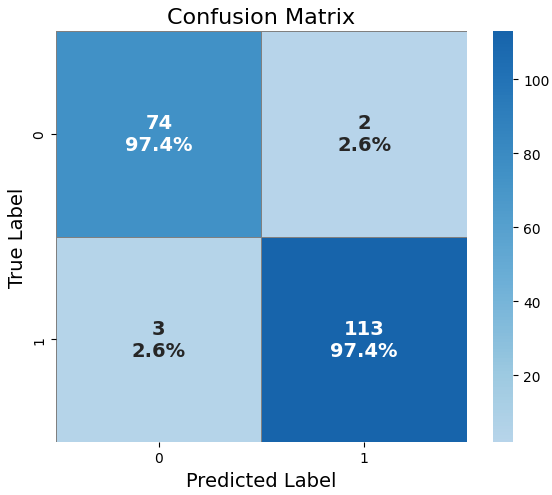}
%		\caption{Visualization of gravitational wave classification results.}
%		\label{fig:fig2-2}
%	\end{subfigure}
	\caption{Identification performance of the finetuned LLM on  LIGO observational data. The model, trained on only 90  events, achieves a recall of 97.4\% for both signal segments and noise segments, and an accuracy of 97.4\%. The misclassification rate of 2.6\% per class demonstrates that the model maintains balanced and reliable performance despite the strongly non-Gaussian and non-stationary noise present in detector data.}
	\label{fig:fig2}
\end{figure}

On the held-out test set, the finetuned model achieves balanced and stable classification performance. As illustrated in Figure~\ref{fig:fig2}, the recall for both the positive class (segments containing gravitational wave signals) and the negative class (noise only segments) reaches 97.4\%, corresponding to a misclassification rate of approximately 2.6\% per class and the entire accuracy of 97.4\%. These results demonstrate that, in the presence of strongly non-Gaussian and non-stationary detector noise, the model maintains high detection efficiency while keeping the false-alarm rate low. The nearly symmetric performance between the two classes further indicates that the finetuning procedure does not bias the model toward any particular outcome. 

This behavior stands in clear contrast to prior work based on traditional neural networks. A common feature of existing studies is their dependence on extensive and carefully constructed simulation sets, with model performance closely tied to the coverage of the simulated parameter space and the fidelity of the injected noise model. Notably, to the best of our knowledge, no previous study has successfully trained a deep neural network for gravitational wave signal detection using only observational data without relying on simulations. 

Taken together, these findings confirm our previous expectation that LLMs are both effective and robust for gravitational wave detections using observational data. It suggests that LLMs may serve as practical and scalable tools for future gravitational wave data analysis, especially in settings where real labeled events are scarce and the noise environment is complex.

\section{Further Investigation} \label{sec:simulation-based-eval}

\subsection{Does Large Simulated Dataset Improve LLM Performance?}

Large scale simulated datasets are commonly used in traditional neural network pipelines because gravitational wave detections are scarce. Previous studies have shown that CNN-based approaches typically rely on hundreds of thousands of simulated injections to obtain stable and reliable performance. Given that our LLM already reaches strong accuracy when trained on only 90 observed events, a further question arises: \textit{does additional simulated data still improve LLM performance?}

To examine this question, we employ the G2Net Gravitational Wave Detection Challenge dataset,\footnote{\href{https://www.kaggle.com/competitions/g2net-gravitational wave-detection/data}{\textit{Kaggle: G2Net Gravitational Wave Detection}}} which consists of simulated compact-binary coalescence signals added to real interferometer noise. The dataset includes 560{,}000 training samples and 226{,}000 test samples, each containing 2 seconds of data from the Hanford, Livingston, and Virgo detectors sampled at 2048 Hz. The task is binary classification of whether a gravitational wave signal is present.

We adopt a two-stage procedure. Starting from the pretrained LLM, we first perform a pre-finetuning phase on the entire simulated dataset to obtain a simulation-adapted model checkpoint. This checkpoint is then finetuned on  observational data following the same preprocessing and training scheme used earlier.

\begin{figure}
	\centering
	% \begin{subfigure}[b]{0.55\textwidth}
	% 	\includegraphics[width=\textwidth]{fig/fig4-1.png}
	% 	\caption{Training loss and validation accuracy of LLaMA3 model.}
	% 	\label{fig:fig4-1}
	% \end{subfigure}
	% \hspace{0.02\textwidth}
	% \begin{subfigure}[b]{0.35\textwidth}
		\includegraphics[width=0.45\textwidth]{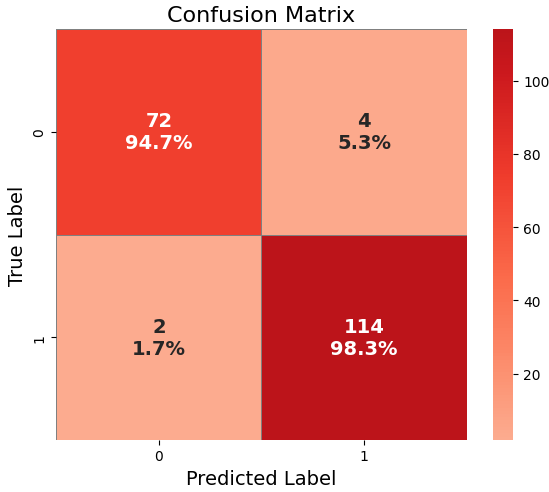}
		% \caption{Visualization of gravitational wave classification results.}
		% \label{fig:fig4-2}
	% \end{subfigure}
	\caption{Confusion matrix for identification on LIGO observational data using a model pre-finetuned on large-scale simulated samples. The performance is comparable to that reported in Figure~\ref{fig:fig2}, indicating that simulation-based pre-finetuning does not provide measurable improvement over finetuning on observational data alone.}

	\label{fig:fig4}
\end{figure}

During the simulation-based pre-finetuning stage, we notice that the training loss decreases very slowly and plateaus at a relatively high value, corresponding to an accuracy of $\sim70\%$. Only after switching to observational data does the loss drop substantially. Evaluation on the held-out observational test set confirms this trend. As shown in Figure~\ref{fig:fig4}, the recall for both signal and noise classes is essentially unchanged relative to models trained solely on observed events. This indicates that the additional 560{,}000 simulated samples do not provide further gains in accuracy.

These findings suggest that, unlike traditional neural networks whose performance strongly depends on large scale simulated datasets, simulation-based pre-finetuning offers limited benefit for basic detection tasks of LLMs, and the discriminative patterns needed for real-data identification appear to be acquired most effectively when the LLMs are exposed directly to observational data.

\subsection{How Does Model Size Improve LLMs Performance?}

\begin{figure}
	\centering
	\includegraphics[width=0.48\textwidth]{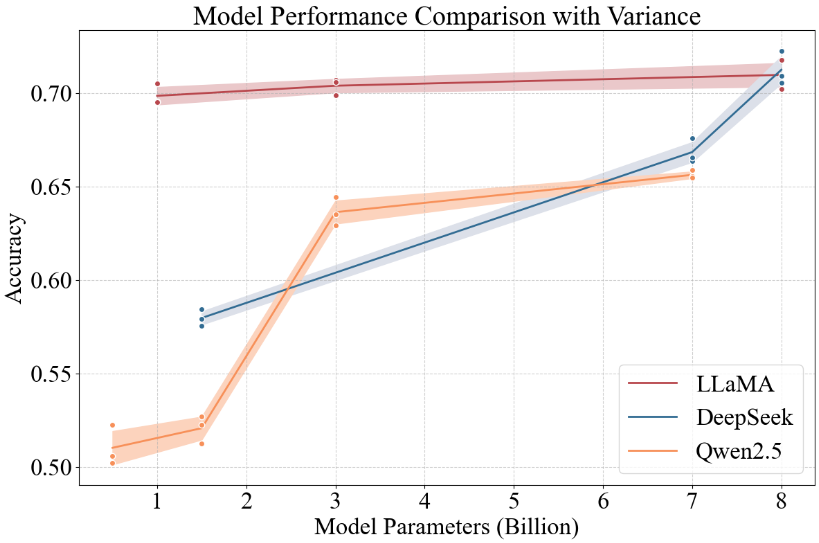}
	\caption{Classification accuracy of LLMs as a function of parameter size. Results are shown for three independent runs of each configuration, the solid line denotes the mean accuracy and the shaded region shows the standard deviation. Accuracy improves steadily as model size increases, and the three architectures converge to similar performance in the 7-8 billion parameter range.}
	\label{fig:fig6}
\end{figure}

Motivated by the scaling law of LLMs \citep{kaplan2020scalinglaws, bubeck2023sparks, 2023arXiv230318223Z}, we investigate how model parameter size affects gravitational wave identification, we evaluate three representative LLM families, Qwen2.5, LLaMA3, and DeepSeek, across a range of parameter sizes from 0.5 billion to 8 billion \citep{2023arXiv230916609B,2024arXiv240721783G,guo2025deepseek}. For fairness, all models use the same dataset of 3{,}000 simulated samples from G2Net and a fixed token cutoff length of 4{,}096. To assess robustness, each configuration is trained and evaluated three times with different random seeds. The resulting accuracies are shown in Figure~\ref{fig:fig6}, where the shaded regions denote the variability across runs. We use simulated data for two reasons. First, the amount of available observational data is possibly too small to reflect the advantage of larger models. Second, we will have a following similar experiment requiring changing the size of the training set, which in turn needs a large pool of samples. Pure observational data cannot meet this requirement.

Across all architectures, accuracy increases with parameter size. LLaMA3 exhibits the strongest small-scale performance, maintaining accuracy above 0.69 even at the 1B scale. DeepSeek and Qwen2.5 improve more gradually, showing substantial gains only once parameters exceed several billion. 

Beyond approximately $\sim8$ billion parameters, the performance of all three model families converges. At this scale, differences between architectures become modest, and all models reach accuracies around 0.70-0.72 with small variance across repeated trials. This convergence suggests that once models reach sufficient parameters, their ability to extract discriminative structure from gravitational wave spectra becomes comparable despite differences in pretraining strategies or architectural design.

Overall, the results show that while larger models consistently produce better performance, the magnitude of improvement depends strongly on architecture in the small-to-mid scale regime. At larger scales, however, the three families achieve similar accuracy, indicating that parameter size eventually dominates architectural differences for this identification task.

%To evaluate performance's efficiency trade-offs, we also employ distilled models from DeepSeek \citep{guo2025deepseek}. Knowledge distillation, widely used in natural language processing to compress large models into smaller architectures while retaining most of their predictive power \citep{2015arXiv150302531H, 2019arXiv191001108S}, is applied here to transfer knowledge from the original 671B DeepSeek-R1 model into compact variants such as DeepSeek-R1-Distill-Llama-8B and 70B. This process yields substantial reductions in memory footprint and inference latency, enabling deployment on more modest hardware. Although the distilled models exhibit a slight decrease in accuracy compared to the full-scale 671B model, their efficiency advantages make them attractive for resource-constrained gravitational wave data analysis.

\subsection{How does Dataset Size Improve LLMs Performance ?}

\begin{figure}
	\centering
	\includegraphics[width=0.48\textwidth]{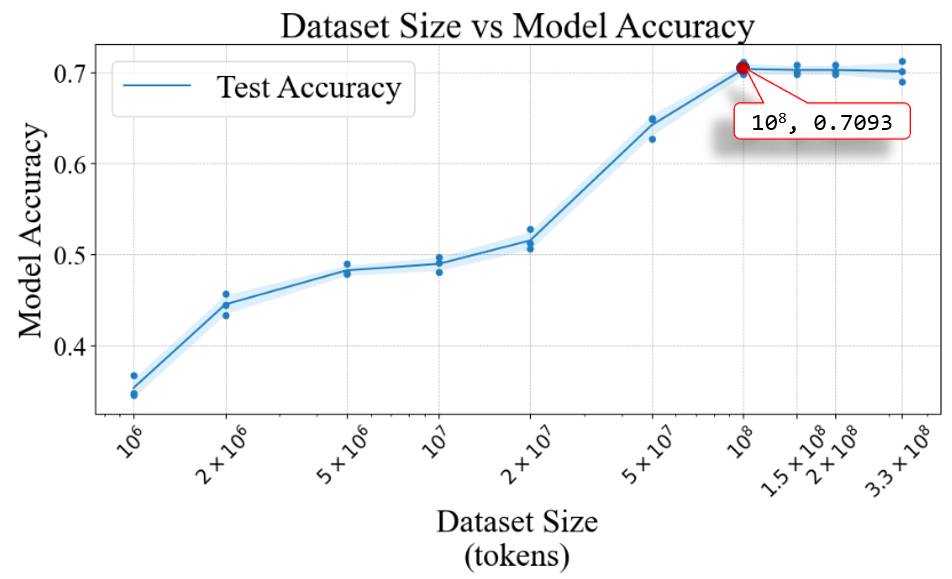}
	\caption{Model accuracy as a function of dataset size.  Each dataset configuration is evaluated over three independent runs, with all individual results shown.  The solid line denotes the mean accuracy and the shaded region shows the standard deviation, illustrating the scalability of LLM performance as training data increases.}
	\label{fig:fig5}
\end{figure}

To examine how the amount of data influences the identification accuracy of large language models, we fine-tune the Llama3 8-billion model on training sets of progressively increasing size. Figure~\ref{fig:fig5} presents the test accuracy obtained after fine-tuning on each dataset configuration. For all runs, the input sequence is truncated to 4096 tokens to ensure consistency. The total size of the training data is controlled by limiting the maximum number of samples, covering a range from 300 to 60{,}000 samples from the simulation, which corresponds to approximately 1.2 million to 240 million tokens.

We observe a clear positive relationship between dataset size and model accuracy. As the amount of data increases, the model exhibits progressively better generalization to the held-out test set. The improvement, however, becomes less steep once the dataset reaches the upper end of our range, indicating that additional data provides smaller incremental gains. This trend is consistent with established scaling behaviors in deep learning, which predict predictable performance improvements as a function of dataset size up to a regime where the benefit begins to slow \citep{2022arXiv220604615S}.

Overall, these results show that while large models such as LLMs can already learn meaningful structure from small observational datasets, additional training data still strengthens performance, particularly in the low data regime. Beyond a certain point, the improvement becomes more modest, suggesting a transition toward a saturation regime.

\section{Applicability to Other Astronomical Domains}\label{sec:other-data}

The analysis presented above suggests that LLMs are particularly effective when three broad conditions are met. First, the discriminative information in the data must be carried primarily by global patterns rather than by local numerical details. In such cases, tokenization and attention naturally emphasize large scale morphology and long range coherence. Second, the noise environment should contain non-Gaussian and non-stationary components, including short duration transients, instrumental artifacts, or drifting spectral features. Local convolutional filters tend to overreact to these outliers, whereas global attention can compare tokens jointly and downweight isolated structures that do not participate in a coherent pattern. Third, labeled examples are often limited in practice. While additional data benefit all models, small sample regimes pose a particular challenge for traditional networks, whereas LLMs can already reach useful accuracy with only 90 examples.

These conditions are not unique to gravitational wave detectors. Several astronomical domains produce time-frequency or time–channel data in which coherent astrophysical structures must be extracted from complex, transient dominated noise. Examples include radio pulsar and fast radio burst searches, dynamic spectra from low frequency interferometers, X-ray timing of accreting neutron stars and black holes, and high energy transient monitors \citep{vaughan2013random,Linares2013The,salvo2021accretion,Hurley-Walker2022A,Dong2022A,Li2023Broadband,Anumarlapudi2025ASKAP}. In many of these settings, the signal of interest forms a smooth or slowly evolving pattern, while the noise sometimes contains short lived, non-stationary, and instrument specific features.

Radio observations provide a concrete example \citep{zhao2023single, yuan2022categorize}. Searches for pulsars and fast radio bursts are commonly performed in dynamic spectra, where intensity is recorded as a function of time and frequency. Astrophysical signals appear as broadband, dispersed sweeps or as repeating pulse trains that are globally coherent across the observing band. In contrast, radio frequency interference (RFI) introduces a wide range of non-Gaussian artifacts, including impulsive bursts, narrowband lines, drifting carriers, and intermittent interference with statistics that vary on short time scales. The noise is highly non-stationary and often dominated by human generated transients rather than by thermal fluctuations. Moreover, labeled examples of rare classes, such as new pulsars or isolated bursts, are scarce relative to the volume of background data. In such a regime, an LLM operating on tokenized patches of the dynamic spectrum could evaluate global consistency across the full time–frequency plane, identify dispersed or repeating structures, and suppress localized RFI that does not conform to any coherent astrophysical pattern. This behavior mirrors the advantages observed in the gravitational wave case and suggests that the methods developed here may transfer naturally to radio astronomy.

\section{Conclusions}\label{sec:conclusions}

This work set out to examine whether LLMs offer practical advantages for processing astronomical data, particularly in regimes where traditional neural networks face limitations. From theoretical considerations, we anticipated that LLMs could be effective when the data are non-Gaussian, non-stationary, and available only in small quantities. Gravitational wave observations provide a natural testing ground for this prediction, as real interferometer data exhibit strong transient noise and only a limited number of confirmed events.

Our real data experiments confirm these expectations. With only 90 observed LIGO events for finetuning and without any assistance from simulated signals, LLMs such as LLaMA3 achieve an identification accuracy of $97.4\%$ on held-out observational samples. The models converge within two epochs, remain stable under non-Gaussian and non-stationary noise, and require no domain-specific pretraining. These results demonstrate that LLMs can extract discriminative structure directly from observational data, even under limited sample size.

We then tested whether additional simulated data still improves LLM performance, a condition under which traditional neural networks typically benefit. Using the large-scale G2Net dataset, we performed a simulation-based pre-finetuning stage before switching to observational data. The effect was negligible: the accuracy on real LIGO data remained essentially unchanged, and the loss during the simulation stage decreased slowly and plateaued well above the levels achieved with real data. This contrasts with conventional CNN pipelines, where simulated datasets are essential for stable performance.

Finally, we conducted scaling experiments. Increasing model size consistently improved accuracy, and models of different families converged to similar performance once they reached the $\sim$8 billion parameter scale. Expanding the training dataset also enhanced accuracy, particularly in the low data regime, with diminishing gains at larger scales. These behaviors align with scaling-law trends observed in other domains.

To conclude, our findings show that LLMs form a viable and efficient alternative to traditional neural networks for gravitational wave signal identification. Their ability to handle complex, non-Gaussian, and non-stationary noise, combined with strong performance in small-data settings, suggests that LLMs may play an important role in future gravitational wave identifications.

\section*{Acknowledgments}

We thank the LIGO Scientific Collaboration for providing access to publicly released gravitational wave datasets. We thank the European Gravitational Observatory (EGO) for providing simulated gravitational wave datasets. We acknowledge support from the National Natural Science Foundation of China (NSFC) under grant No. 12588202 (PI: Di Li). We are grateful to Pei Wang (National Astronomical Observatories, CAS), and to Dongzi Li, Lei Zhang, and Yongkun Zhang (Tsinghua University) for helpful discussions that improved this work.

\bibliography{refs}
\bibliographystyle{aasjournal}



\end{document}